\documentclass[fleqn,twoside]{article}
\usepackage{espcrc2}

\newcommand{\be}{\begin{equation}}
\newcommand{\ee}{\end{equation}}

\newcommand{\pslash}{\hbox{$p$\kern-0.5em\raise-0.35ex\hbox{$/$}}\thinspace}

\newcommand\FF{{\cal F}}
\def\sfrac#1#2{{\textstyle\frac{#1}{#2}}}

\title{High-accuracy two-loop computation of the critical mass for 
Wilson fermions }

\author{Sergio Caracciolo\address{Dipartimento 
         di Fisica, Universit\`a degli Studi di Milano, 
         via Celoria 16, I-20133 Milano, 
         INFN, Sezione di Pisa and NEST-INFM, Italy},
        Andrea Pelissetto\address{Dipartimento di Fisica and INFN --
         Sezione di Roma I, 
         Universit\`a degli Studi di Roma ``La Sapienza", 
         P.le Aldo Moro 2, I-00185 Roma, 
         Italy
        }, and
        Antonio Rago\address{Dipartimento di Fisica and INFN --
         Sezione di Torino,
        Universit\`a degli Studi di Torino, \\
        Via Pietro Giuria 1, I-10125 Torino, Italy
        }}
       
\begin{document}

\begin{abstract}
We test an algebraic algorithm based on the 
coordinate-space method,  evaluating with high accuracy
the critical mass for 
Wilson fermions in lattice QCD at two loops. We test the results by using 
different types of infrared regularization.  
 
\vspace{1pc}
\end{abstract}

\maketitle

We have already presented at a Lattice conference~\cite{Capitani} an 
algebraic 
algorithm that allows to apply the coordinate-space method  by
L\"uscher and Weisz~\cite{LW} to two-loop lattice integrals with gluon and 
Wilson-fermion propagators. In order to test the method, 
we have recently redone \cite{Rago} a two-loop computation  of the
critical mass for Wilson fermions~\cite{Follana}.

For Wilson fermions,
the dressed inverse fermion propagator has the form
\begin{eqnarray}
\lefteqn{S^{-1}(p,m_B) =}\nonumber \\
& =\, i\,\overline{\pslash}\, +  m_B + M_W(p)
- \Sigma^L(p,m_B,g_0), 
\end{eqnarray} 
where, setting the lattice
spacing equal to one,
\begin{eqnarray}
\overline{p}_\mu &=&  \sin p_\mu \; ,\\
 \hat{p}^2 &=& \sum_\mu \left( 2 \sin {p_\mu \over 2} \right)^2 \; , \\
 M_W(p) &=& {r_W\over2} \hat{p}^2 \; .
\end{eqnarray}
The additive mass renormalization $\delta m_B$ 
is obtained by requiring 
$S^{-1}(0,\delta m_B) =0$, i.e.
\be
\Sigma^L(0,\delta m_B,g_0) = \delta m_B.
\ee
This equation can be solved in perturbation theory 
by expanding 
\be
\Sigma^L(0,m_B,g_0) = \sum_{n=1}^\infty g_0^{2n} \,\Sigma^{(n)}.
\ee 
We have computed
$\Sigma^{(1)}$ and $\Sigma^{(2)}$ for $r_W=1$, 
gauge group $SU(N)$, and $N_f$ fermionic flavour species, in the Feynman 
gauge.

In Ref.~\cite{Burgio} we already reported the analytic
one-loop expression for the fermionic self-energy $\Sigma^L$. 
Our result was expressed in
terms of three purely bosonic constants $Z_0$, $Z_1$ and $F_0$ and
of 12 numerical constants that appear in the presence of Wilson
fermions.  
The numerical values of these constant are obtained by using a
powerful recursive method that gives very precise 
results~\cite{Burgio,Burgio:lat}.
This algorithm generalizes the method we introduced for purely bosonic 
integrals in~\cite{Menotti}.
In practice, we have computed all 
constants but $F_0$ with 60-digit precision.

At one-loop order 
\be
\Sigma^{(1)} = {N^2-1 \over N}\, \sum_{i=1}^2 c_{i}^{(1)},
\ee     
where $c_{i}^{(1)}$ are the contributions of the two contributing diagrams. 
In terms of the basic integrals they are given by 
\begin{eqnarray*}
c^{(1)}_1 & = &  - Z_0\\
c^{(1)}_2 & = &  {Z_0 \over 2} -  \,\FF(1,0)\; .
\end{eqnarray*}
Summing up the two contributions we obtain
\begin{eqnarray}
\sum_{i=1}^2 c_{i}^{(1)} & = & - \left[ Z_0 + 2 \,\FF(1,0)
\right] \nonumber \\&  \approx &- 0.16285705871085078618\; .
\end{eqnarray}
The constant is in excellent agreement with the result 
of Ref.~\cite{Follana},\par 
\noindent $ \sum_i c_{i}^{(1)} = - 0.162857058711(2)$.

At two loops there are 26 diagrams. They are numbered as 
in Ref.~\cite{Follana} in order to simplify the comparison. 
The $i$-th diagram gives a contribution of the form
\be
D_i \equiv (N^2-1) \left[ c_{1,i}^{(2)} + {1\over N^2} \, c_{2,i}^{(2)} +
{N_f\over N} \, c_{3,i}^{(2)} \right].
 \ee
In Table~\ref{compare} we report the results given in
Ref.~\cite{Follana} and those obtained here by using the configuration-space 
method.
\begin{table*}[htb]
\begin{center}
\protect\tiny
\begin{tabular}{cr@{}lr@{}lr@{}l}
\hline
\hline 
$i$&\multicolumn{2}{c}{$c^{(2)}_{1,i}$}&\multicolumn{2}{c}{$c^{(2)}_{2,i}$}&
\multicolumn{2}{c}{$c^{(2)}_{3,i}$}\\  
\hline
\hline 
3 & 0&.002000362950707492 & -0&.0030005444260612375& 0&\\ 
& 0&.0020003629507074987148 & $-$0&.0030005444260612480722 & 0& \\ 
\hline 
4 & 0&.00040921361(1) & $-$0&.00061382041(2) & 0& \\ 
& 0&.0004092136068803147865 & $-$0&.0006138204103204721798 & 0&\\ 
\hline 
5 & 0& & 0& & 0& \\ 
& 0& & 0& & 0&\\ 
\hline 
6 & $-$0&.0000488891(8) & 0&.000097778(2) & 0& \\ 
& $-$0&.0000488853119(2)& 0&.000097770623(5)& 0&\\ 
\hline 
7+8+9+10+11 & $-$0&.013927(3) & 0&.014525(2) & 0& \\ 
& $-$0&.01392647740 (2)& 0&.0145250053341618950704 & 0&\\ 
\hline 
12+13 & 0& & 0& & 0&.00079263(8) \\ 
& 0& & 0& & 0&.000792647(2)\\ 
\hline 
14+15+16+17+18 & $-$0&.005753(1) & 0&.0058323(7) & 0& \\ 
& $-$0&.00575248584(1) & 0&.005832127004694453 & 0&\\ 
\hline 
19+20 & 0& & 0& & 0&.000393556(7) \\ 
& 0& & 0& & 0&.000393556(4)\\ 
\hline 
21+22+23 & 0&.000096768(4) & $-$0&.000096768(4) & 0& \\ 
& 0&.0000967648(2) & $-$0&.0000967648(2) & 0&\\ 
\hline 
24 & 0& & 0& & 0& \\ 
& 0& & 0& & 0&\\ 
\hline 
25 & 0&.00007762(1) & $-$0&.00015524(3) & 0& \\ 
& 0&.000077613106(4)& $-$0&.000155226212(8)& 0& \\ 
\hline 
26 & $-$0&.00040000(5) & 0& & 0& \\ 
& $-$0&.00039997586(1)& 0& & 0&\\ 
\hline 
27 & 0& & $-$0&.000006522(1) & 0& \\ 
& 0& & $-$0&.0000065203(1)& 0& \\  
\hline 
28 &  0&.0000078482(5) & $-$0&.000015696(1) & 0& \\ 
& 0&.0000078480652722033294 & $-$0&.0000156961305444066589 & 0& \\ 
\hline 
\hline 
Total & $-$0&.017537(3) & 0&.016567(2) & 0&.00118618(8) \\ 
& $-$0&.0175360218(2) & 0&.0165663304(2) & 0&.001186203(6)\\ 
\hline
\hline
\end{tabular}
\end{center}
\caption{Coefficients $c^{(2)}_{1,i}$, $c^{(2)}_{2,i}$ and
$c^{(2)}_{3,i}$.  For each of them
we report in the first line the result of Ref.~\cite{Follana},
obtained by means of a momentum-space integration, 
and in the second line our result, obtained by means of 
the coordinate-space method.}
\label{compare}
\end{table*}
When we have not reported an error, the precision we achieve is much
higher than the reported digits. This occurs in general when the 
diagram is the product of one-loop integrals.  
All results are in agreement with those presented
in Ref.~\cite{Follana}. Only for diagram 6 there is apparently
a (very) small underestimation of the error, 
which is negligible in the sum of all contributions.
In Table~\ref{compare} diagrams are grouped together
in order to obtain infrared-convergent results, but this not
necessary in our method. 
Indeed, we can compute each of them separately, by introducing an infrared 
regularization.
To test the results, we have used four different infrared regularizations:
\begin{itemize}
\item[(a)] We add a mass in the denominators of the
  propagators. Explicitly, for  
the gluon ($\Delta_B(k)$) and for the fermion ($\Delta_F(k)$) propagator
we use:
\begin{eqnarray*}
\Delta_B(k) &=& \frac{1}{ \hat{k}^2 + m^2}\\
\Delta_F(k) &=& \frac{- i \bar{k}_\mu \gamma_\mu +
  M_W(k)}{\overline{k}^2 + M_W(k)^2  + m^2} 
\end{eqnarray*}
\item[(b)] We regularize the gluon propagator as in (a), but
  use instead the correct Wilson-fermion propagator
\begin{eqnarray*}
\Delta_F(k) = \frac{- i \bar{k}_\mu \gamma_\mu + M_W(k) + m} 
                {\bar{k}^2 + (M_W(k) + m)^2}
\end{eqnarray*}
\item[(c)] We regularize the Wilson fermion as in (a), but use the 
massless propagator for the gluon.
\item[(d)] We regularize the Wilson fermion as in (b) and the 
massless propagator for the gluon.
\end{itemize}
As an example, we report in Table~\ref{2} the divergent and the finite
contribution   of three diagrams whose sum is infrared finite.
\begin{table*}[htb]
\begin{center}
\begin{tabular}{c l l l l}
\hline
\hline
$i$ & Divergent Part & $c_{1,i}^{(2)}$ & $c_{2,i}^{(2)}$ & $c_{3,i}^{(2)}$\\
\hline
\hline
&&&&\\[-1mm]
21 & $\left(1-\sfrac{1}{N^2}\right)\sfrac{Z_0}{8\pi^2}\log m^2$ &
0.001606284825541242 & --0.001606284825541242 &0\\[1mm] 
22 & $\left(1-\sfrac{1}{N^2}\right)\left(\sfrac{-Z_0}{16\pi^2} +
  \sfrac{{\cal F}(1, 0)}{8\pi^2}\right)\log m^2$ & 0.0005015205(2) &
--0.0005015205(2)&0\\[1mm] 
23 & $\left(1-\sfrac{1}{N^2}\right)\left(\sfrac{-Z_0}{16\pi^2} -
  \sfrac{{\cal F}(1, 0)}{8\pi^2}\right)\log m^2$& --0.002011040454066014 &
0.002011040454066014&0\\[1mm] 
\hline
\hline
\end{tabular}
\end{center}
\caption{Divergent and finite contribution 
of three diagrams whose sum is infrared finite.}\label{2}
\end{table*}
If we write
$$D_i^{(x)} = D_i^{(a)} + \left(1 - {1\over N^2}\right) \Delta^{(x)}_i,
$$ 
where $(x)$ refers to the chosen infrared regularization, we get the
results
reported in Table~\ref{3}.
\begin{table*}[htb]
\begin{center} 
\begin{tabular}{clll} 
\hline 
\hline 
$i$ &  $\Delta^{(b)}_i$ & $\Delta^{(c)}_i$ & $\Delta^{(d)}_i$
\\ 
\hline 
\hline 
&\\[-3mm] 
21 & $+\sfrac{Z_0}{16\pi^2}$ & $+\sfrac{11 Z_0}{96 \pi^2} $ & $+\sfrac{16
  Z_0}{96 \pi^2} $ \\[1mm] 
22 & $-\sfrac{Z_0}{32\pi^2} + 
\sfrac{{\cal F}(1,0)}{16\pi^2}$ &$-\sfrac{11 Z_0}{192 \pi^2} +
\sfrac{11 {\cal F}(1,0)}{96 
    \pi^2}$ & $-\sfrac{16 Z_0}{192 \pi^2} + \sfrac{16 {\cal F}(1,0)}{96
    \pi^2}$\\[1mm] 
23 & $-\sfrac{Z_0}{32\pi^2} - \sfrac{{\cal F}(1,0)}{16\pi^2}$ & 
$ -\sfrac{11 Z_0}{192 \pi^2} - \sfrac{11 {\cal F}(1,0)}{96
    \pi^2}$ & $-\sfrac{16 Z_0}{192 \pi^2} - \sfrac{16 {\cal F}(1,0)}{96
    \pi^2}$\\[1mm] 
\hline 
\hline 
\end{tabular} 
\end{center}
\caption{Contribution from different regularizations.}
\label{3}
\end{table*}
Individual diagrams depend on the regularization but their sum does not.

\end{document}